\documentclass[journal=aamick,manuscript=article, layout=onecolumn]{achemso}

\usepackage[version=3]{mhchem} % Formula subscripts using \ce{}
\usepackage{xr}
\usepackage{xcolor, soul}
\sethlcolor{white}
\usepackage{lineno}
\author{Shidong Li}
	\email{li_shidong@ices.a-star.edu.sg/lisd_pe@hotmail.com}
	\affiliation{Institute of Chemical and Engineering Sciences (ICES), Agency for Science,         Technology and Research (A*STAR), 1, Pesek Road, Jurong Island, Singapore, 627833}
\author{Anqi Sng}
    \affiliation{Institute of Materials Research and Engineering, Agency for Science, Technology     and Research (A*STAR), 2 Fusionopolis Way, Innovis, Singapore 138634}
\author{Dan Daniel}
	\email{daniel@imre.a-star.edu.sg}
	\affiliation{Institute of Materials Research and Engineering, Agency for Science, Technology     and Research (A*STAR), 2 Fusionopolis Way, Innovis, Singapore 138634}
\author{Hon Chung Lau}	
    \affiliation{$^{3}$Department of Civil and Environmental Engineering, National University of     Singapore, 1 Engineering Drive 2, Singapore, 117576}
\author{Ole Tors$\ae$ter}
    \affiliation{PoreLab, Norwegian Center of Excellence, S. P. Andersens vei 15b, Trondheim,     Norway, 7031}
    \altaffiliation{Department of Geoscience and Petroleum, Norwegian University of Science and         Technology (NTNU), S.P. Andersens veg 15a, Trondheim, Norway, 7031 }
\author{Ludger P. Stubbs}
    \affiliation{Institute of Chemical and Engineering Sciences (ICES), Agency for Science,         Technology and Research (A*STAR), 1, Pesek Road, Jurong Island, Singapore, 627833}

\title{Visualizing and Quantifying Wettability Alteration 
by Silica Nanofluids} 


\begin{document}
\linenumbers
\begin{tocentry}

\includegraphics[scale=1.0]{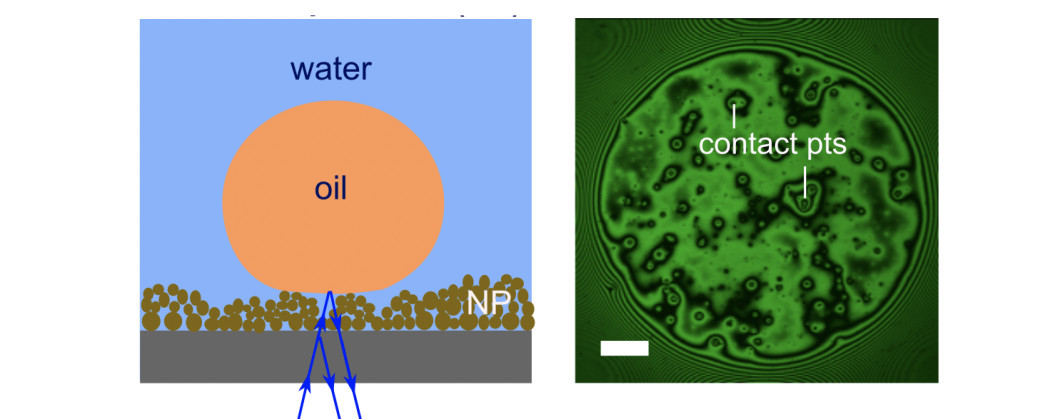}
%Some journals require a graphical entry for the Table of Contents.
%This should be laid out ``print ready'' so that the sizing of the
%Inside the \texttt{tocentry} environment, the font used is Helvetica
%8\,pt, as required by \emph{Journal of the American Chemical
%The surrounding frame is 9\,cm by 3.5\,cm, which is the maximum
%permitted for  \emph{Journal of the American Chemical Society}
%graphical table of content entries. The box will not resize if the
%content is too big: instead it will overflow the edge of the box.
%
%This box and the associated title will always be printed on a
%separate page at the end of the document.

\end{tocentry}

\begin{abstract}
An aqueous suspension of silica nanoparticles or nanofluid can alter the wettability of surfaces, specifically by making them hydrophilic and oil-repellent under water. Wettability alteration by nanofluids have important technological applications, including for enhanced oil recovery and heat transfer processes. A common way to characterize the wettability alteration is by measuring the contact angles of an oil droplet with and without nanoparticles. While easy to perform, contact angle measurements do not fully capture the wettability changes to the surface. Here, we employed several complementary techniques, such as cryo-scanning electron microscopy, confocal fluorescence and reflection interference contrast microscopy and droplet probe atomic force Microscopy (AFM), to visualize and quantify the wettability alterations by fumed silica nanoparticles. We found that nanoparticles adsorbed onto glass surfaces to form a porous layer with hierarchical micro- and nano-structures. The porous layer is able to trap a thin water film, which reduces contact between the oil droplet and the solid substrate. As a result, even a small addition of nanoparticles (0.1 wt$\%$) lowers the adhesion force for a 20-$\mu$m-sized oil droplet by more than 400 times from 210$\pm$10 nN to 0.5$\pm$0.3 nN as measured using droplet probe AFM. Finally, we show that silica nanofluids can improve oil recovery rates by 8$\%$ in a micromodel with glass channels that resemble a physical rock network.       
\end{abstract}

\section*{Introduction}

Suspensions of nanoparticles or nanofluids have many important technological applications, including for enhanced heat transfer and oil recovery processes \cite{taylor2013small, eltoum2020effect}. The wetting properties of surfaces are altered when nanoparticles adsorb onto the surface \cite{wasan2003spreading}; for example, with silica and other metal oxide nanoparticles, the surface becomes more water-wettable or hydrophilic. This greatly improves the heat transfer rates during boiling by preventing air bubbles (which conduct heat poorly) from accummulating on the surface \cite{coursey2008nanofluid}. In enhanced oil recovery (EOR) processes, nanoparticles allow oil trapped in the porous network to be more easily extracted by water injection \cite{kong2010applications, lau2017nanotechnology, li2013improved}.  

The wettability of oil reservoirs plays a critical role in EOR. It can affect the oil-water relative permeability, capillary pressure and therefore the residual oil distribution. Suitably altering reservoir’s wettability, such as using nanofluids, can increase oil recovery. For example, nanoparticles can change the wettability of a carbonate reservoirs from hydrophobic to hydrophilic, improving oil recovery \cite{nazari2015comparative, dehghan2016potential}. The wettability alteration mechanisms of nanofluid have been investigated previously. They include structural disjoining pressure \cite{nazari2015comparative}, increase of surface roughness, reduction of surface free energy \cite{ni2018synthesis}, electrostatic repulsion promotion, decline of non-electrostatic adhesion force and structural interaction force \cite{afekare2021enhancing}, nanoparticle layering near the solid substrate \cite{lim2015nanofluids}, and partial release of stearate from the calcite surface and replacement by silica nanoparticles \cite{dehghan2016potential}. 

\begin{figure}[!htb]
\centering
\includegraphics[scale=1.0]{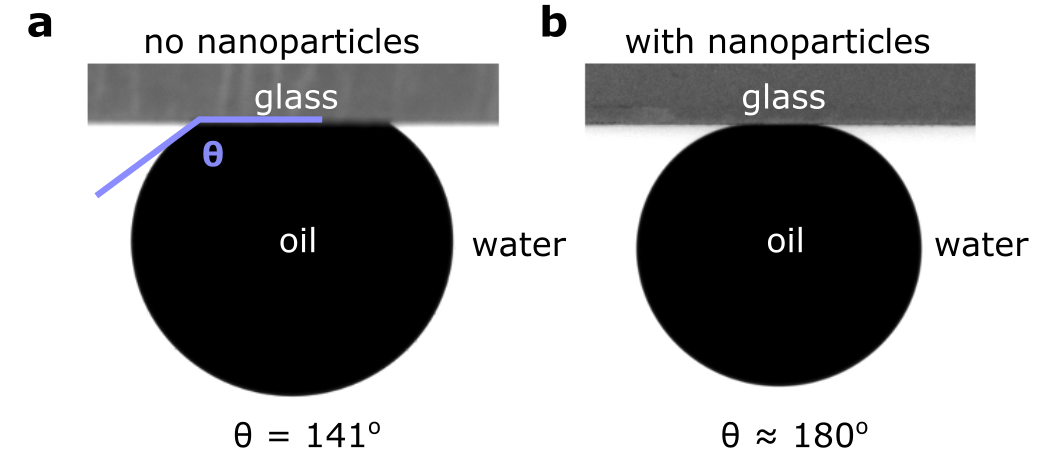}
\caption{\label{fig:angle} Underwater contact angle of crude oil droplets in the absence and presence of nanoparticles.}
\end{figure}

One method to quantify surface wettability alteration is by measuring the contact angles $\theta$ made by air bubbles (for boiling) and oil droplets (for oil recovery processes) under water. For example, adding 0.5 wt$\%$ fumed silica nanoparticles increases $\theta$ from 141$^{\circ}$ to nearly 180$^{\circ}$ for millimetre-sized crude oil droplet (Fig.~\ref{fig:angle}); the glass substrate has become more hydrophilic. For large oil droplets on flat surfaces, the contact angles can be deduced optically from high-resolution sideview images of the droplets as shown in Figure \ref{fig:angle}. For oil trapped inside a porous network, the contact angles at the pore-scale can be measured from X-ray micro-computed tomography (micro-CT) images \cite{andrew2014pore, mirchi2018pore}, which makes this technique particularly suited to understand oil-recovery processes. While powerful and easy to perform, contact angle measurements suffer from several disadvantages  \cite{decker1999physics, schellenberger2016water}. Contact angle measurements are inaccurate for large values close to 180$^{\circ}$ (including in Fig.~\ref{fig:angle}b); a small error in the positioning of the droplet base (as small as a single pixel) translates to  a large error in the contact angle value (of more than $10^{\circ}$) \cite{Liu1147, srinivasan2011assessing}. More importantly, the sideview droplet images in contact angle measurements do not allow us to visualize the three-phase contact line which is better captured from the bottom of the droplet. 
        
Previously, AFM with a sharp solid tip has been used to probe properties of rock surfaces relevant to the oil industry, including their topography, mechanical and chemical properties \cite{javadpour2012atomic, yang2017nanoscale}. The AFM tip can also be functionalized with different organic groups (chemical force microscopy) to mimic the interactions between different oil components and the solid substrate \cite{hilner2015effect,  afekare2020enhancing, afekare2020insights, afekare2021enhancing}. Intermolecular forces and the adhesion between the chemically-modified tips and substrate can be measured with piconewton resolutions, and wettability alterations can be resolved spatially with submicron resolutions using this technique. Chemical force microscopy with its solid tip cannot however fully capture the interactions of a liquid droplet with a solid substrate.

%, in particular the dynamics of contact line pinning and depinning of an oil droplet.  
Here, we propose two techniques, namely reflection interference contrast microscopy (RICM) and droplet probe AFM, that are highly complementary to contact angle measurement and chemical force microscopy, respectively. RICM allows us to observe the droplet's base as opposed to sideview images in contact angle measurements, while droplet probe AFM allows us to measure the interaction forces of a microdroplet (as opposed to a solid tip) with the substrate. 

%This allows us to observe pinning and depinning dynamics of the droplet's contact line, which is not possible with the solid tip in chemical force microscopy. 
We used the two techniques to visualize and quantify wettability alterations by fumed silica nanoparticles, chosen for their low cost and wide availability. Our previous studies also demonstrated the effectiveness of fumed silica in EOR: the injection of 0.05 wt$\%$ fumed silica suspension improves oil recovery from a core plug by 6$\%$ \cite{hendraningrat2013coreflood}.  First, we used cryo-scanning electron microscopy (cryo-SEM) and atomic force microscopy (AFM) to visualize and measured the thickness of adsorbed nanoparticles on a glass surface. We found that the nanoparticles form a porous layer with hierarchical micro- and nano-structures that stabilizes a thin water film beneath the oil droplet, which we confirmed using a combination of confocal fluorescence and RICM \cite{limozin2009quantitative, daniel2017oleoplaning}. With RICM, we show that the oil droplet is in contact only with the topmost tips of the porous layer (Cassie-Baxter state) \cite{quere2008wetting}; as a result, the oil droplet has a high underwater contact angle $\theta$ close to 180$^{\circ}$ with minimal adhesion, reminiscent of the underwater superoleophobicity exhibited by fish scales and artificial bioinspired surfaces \cite{liu2009bioinspired}. We also use droplet probe AFM to directly measure the interaction forces (with  a resolution of less than 1 nN) between an oil microdroplet and a glass surface under different nanoparticle concentrations \cite{daniel2019mapping, daniel2020quantifying, xie2017surface}. We show that fumed silica can reduce the adhesion force and the energy required to remove oil microdroplets by more than 400 and 7000 times, respectively. We also observed force jumps in the force curves corresponding to pinning and depinning dynamics of the contact line. 

When combined together, the techniques described in this paper are able to probe changes in the surface structural and wetting properties by nanofluids in unprecedented details, and highly complement previously described techniques. The techniques presented allow us to better understand an important aspect of how wettability alteration is achieved by nanofluids: the adsorbed nanoparticles are able to stabilize a thin water film beneath the oil droplet. With RICM, we are able to visualize this water film and the pinning points directly, while droplet probe AFM allows us to measure the adhesion due to the pinning points, including the dynamics of contact line pinning-depinning. Finally, we demonstrate the effect of wettability alteration on the effectiveness of oil recovery in a micromodel with glass channels that resemble a physical rock network and observed an improvement in oil recovery rate of 8$\%$.

\section*{Results and discussion}

\subsection*{Structure of adsorbed nanoparticles}

\begin{figure}[!htb]
\centering
\includegraphics[scale=1.0]{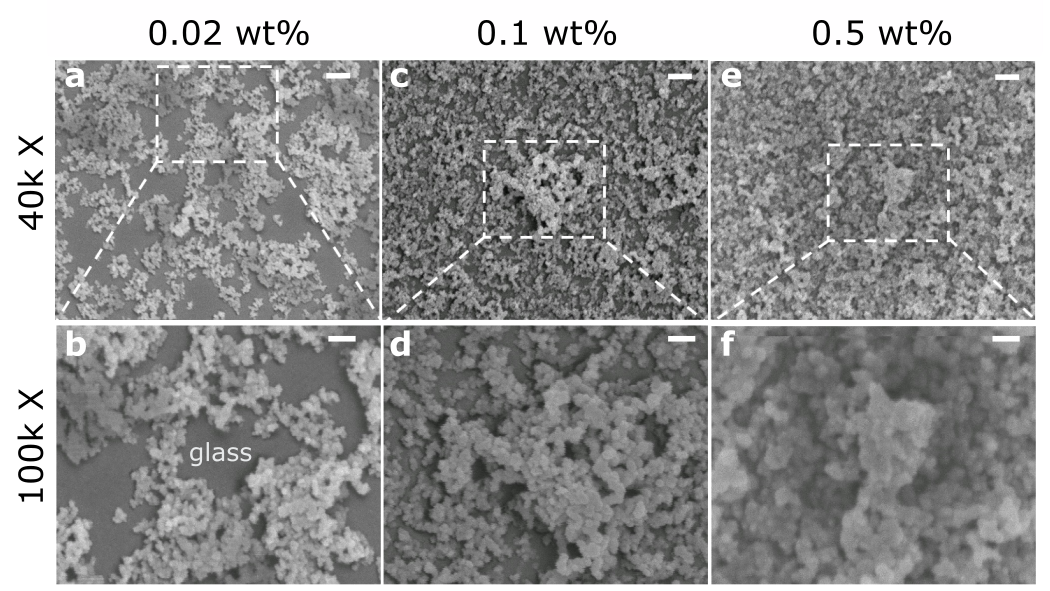}
\caption{\label{fig:SEM} Cryo-SEM of adsorbed nanoparticles at different particle concentrations 0.02--0.5 wt\% and under 40,000$\times$ and 100,000$\times$ magnifications. Scale bars are 200 nm for (a, c, e) and 100 nm for (b, d, f).}
\end{figure}

\begin{figure}[!htb]
\centering
\includegraphics[scale=1.0]{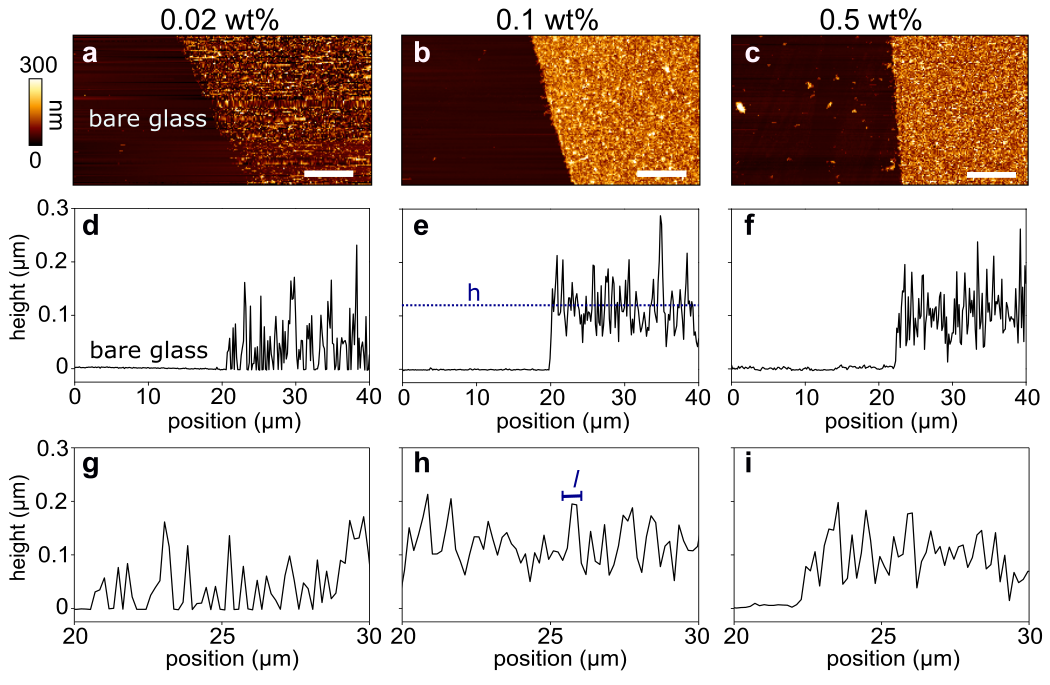}
\caption{\label{fig:AFM} (a--c) The thickness and the roughness of the nanoparticle layer can be determined using atomic force microscopy for different particle concentrations. Scale bars are 5 $\mu$m. (d--f) Cross-sections of the nanoparticle layer for the different particle concentrations and (g--i) zoomed-in views of the cross-sections. }
\end{figure}

\newpage

We first used cryo-SEM to visualize the adsorbed nanoparticles (average primary particle size $s$ is 7 nm) on a glass surface, after it is immersed in brine solution (3 wt$\%$ NaCl in water) containing different particle concentrations from 0.02 to 0.5 wt\% for an hour. 3 wt$\%$ NaCl is used because this is close to the salt content found in seawater which is widely used in EOR applications. In DI water, both glass and silica surface are negatively charged, and in the absence of any salt, the electric double layer forces are sufficiently repulsive to keep the nanoparticles in solution \cite{israelachvili2011intermolecular}. The addition of NaCl salt can however screen the double layer forces and promotes the aggregation and adsorption of nanoparticles; as a result, the nanoparticles form porous structures with hierarchical micro- and nano-sized features on glass surfaces \cite{lin1989universality}. At the lowest particle concentration of 0.02 wt$\%$, the surface is only partially covered with nanoparticles and bare glass is visible on the cryo-SEM image (Fig.~\ref{fig:SEM}a, b); in contrast, there is full coverage of nanoparticles above 0.1 wt$\%$ particle concentration (Fig.~\ref{fig:SEM}c--f).  Note that before imaging in cryo-SEM, the surface is flash frozen using liquid nitrogen while still wet to preserve the original structures of the nanoparticle aggregates.  

The thickness of the nanoparticle layer $h$ can be determined by performing AFM at the boundary between  the adsorbed layer and bare glass (Figs.~\ref{fig:AFM}a--c, See Materials and Methods for experimental details). Experimentally, we found that $h$ = 47$\pm$46, 120$\pm$35, 108$\pm$40 nm for 0.02, 0.1 and 0.5 wt$\%$ particle concentrations, respectively (Figs.~\ref{fig:AFM}d--f). The reported errors in $h$ are the arithmetic mean roughness $R_a$. 

Since the nanoparticle aggregates have a typical lateral dimension $l$ of about 1 micron (Figs.~\ref{fig:AFM}g--i), we can estimate the number of particles in each aggregate $N \sim hl^{2}/s^{3} \approx 10^{5}$. Note that there may be aggregates with features larger than reported here but not be picked up because of the inherently small scan range of the AFM technique (typically less than 0.1 mm).  

As will be discussed in the next sections, the presence of this nanoparticle layer with hierarchical micro- and nano-sized features have important consequences for the surface wetting poperties.

\subsection*{Water film stabilized by nanofluid}

\begin{figure}[!htb]
\centering
\includegraphics[scale=0.9]{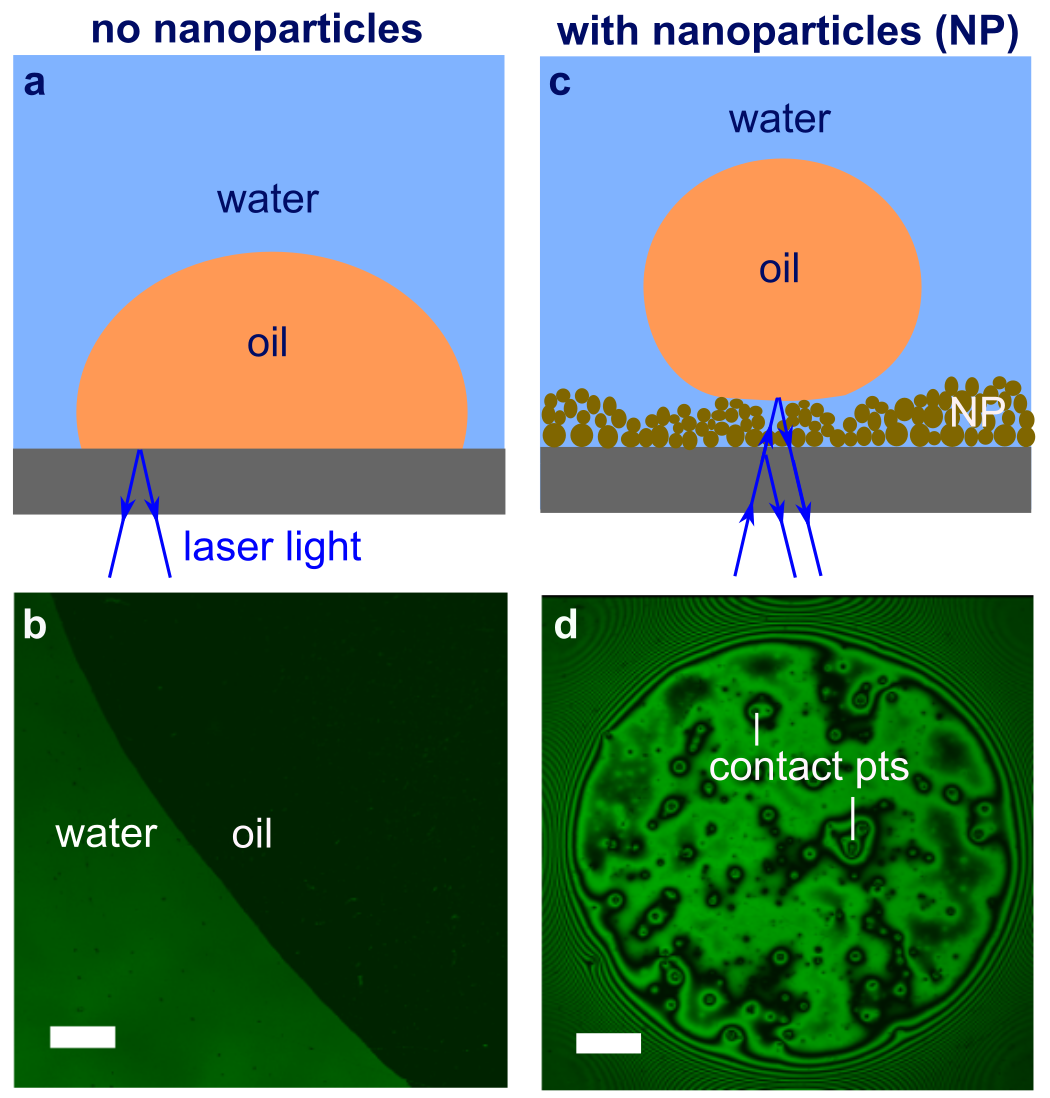}
\caption{\label{fig:base} (a, b) Schematic and micrograph showing the oil droplet in contact with the glass substrate with no nanoparticles. Scale bar is 100$\mu$m. (c, d) The nanoparticles are able to stabilize a thin water film beneath the oil droplet and the droplet is in contact only with the topmost tips of the nanoparticle aggregates. Scale bar is 100 $\mu$m.}
\end{figure}

\begin{figure*}[!htb]
\centering
\includegraphics[scale=1.4]{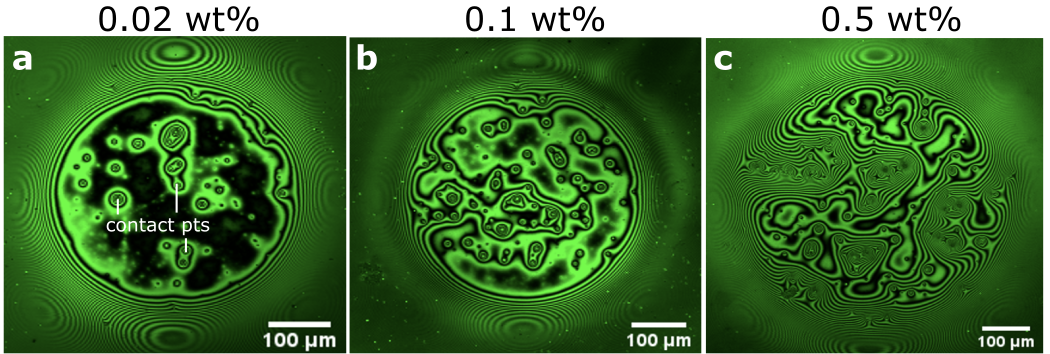}
\caption{\label{fig:diff_conc} Droplet's base as observed using RICM for different particle concentrations. Scale bars are 100 $\mu$m. }
\end{figure*}

In the absence of nanoparticles, a millimetric-sized oil droplet is in contact with the underlying glass substrate when submerged under water, as shown in the schematic of Fig.~\ref{fig:base}a. The micrograph in Fig.~\ref{fig:base}b, taken using confocal reflection interference contrast microscopy (RICM)\cite{limozin2009quantitative, daniel2017oleoplaning}, shows the edge of the droplet's base at the three-phase contact line. In confocal RICM, the surface is raster scanned with a focused beam of monochromatic light  (wavelength $\gamma$=561 nm) and the intensity of the reflected signal $R$ increases with increasing refractive index contrast $\Delta n$: $R \propto \Delta n^{2}$ \cite{born2013principles}. Hence, the droplet's base appears much darker than the surrounding, because of the smaller $\Delta n$ between glass and the silicone oil chosen ($n_{\text{glass}}$ = 1.52, $n_{\text{oil}}$ = 1.51, $\Delta n$ = 0.01) compared to $\Delta n$ between glass and water ($n_{\text{water}}$ = 1.33, $\Delta n$ = 0.19). Details of the technique can be found in our previous publication \cite{daniel2017oleoplaning}.  

In contrast, there is no direct contact between the same silicone oil droplet ($\sim$ 4 mm diameter) and the underlying substrate with the addition of 0.1 wt$\%$ nanoparticles (Fig.~\ref{fig:base}c). The micrograph in Fig.~\ref{fig:base}d shows the droplet's base ($\sim$ 0.5 mm in size) which appear as a circular patch; the adsorbed particle layer stabilizes a thin water film beneath the droplet, which result in bright and dark fringes as lights reflected off the glass-water and water-oil interfaces interefere constructively or destructively depending on the local film thickness. The difference in water film thickness $\Delta h_{\text{water}}$ between adjacent bright and dark fringes is $\sim \lambda/4n_{\text{water}} \approx 200$ nm; at 0.1 wt$\%$ particle concentration, the variation in water film thickness ($\sim$ 200 nm) is small compared to the size of the droplet's base ($\sim$ 0.5 mm).   

The water film can be stabilized with as little as 0.02 wt$\%$ nanoparticle concentration (Fig.~\ref{fig:diff_conc}a), even though the nanoparticle surface coverage at this concentration is only partial (See Fig.~\ref{fig:SEM}). As we increase the particle concentration to 0.5 wt$\%$ (Fig.~\ref{fig:diff_conc}c), the water film thickness under the droplet is no longer uniform, as the droplet's base conform to the shape of the larger aggregates. RICM therefore allows us to observe the droplet's base directly, including the three-phase contact line and pinning points, not easily visualized from the sideview images in contact angle measurements. Here, we show that the oil droplet is in contact only with the topmost tips of the aggregates, i.e. Cassie-Baxter state, which appear as dots surrounded by interference rings on the micrograph (Fig.~\ref{fig:base}d and Fig.~\ref{fig:diff_conc}a). As will be shown in the next section, each of these contacts will act discrete pinning points which can be detected and quantified using droplet probe AFM. Note that the RICM images were taken 15 minutes after the oil droplet was deposited and no more changes to the droplet's base and contact line were observed.

\subsection*{Reduction in oil droplet adhesion}

\begin{figure*}[!htb]
\centering
\includegraphics[scale=1.05]{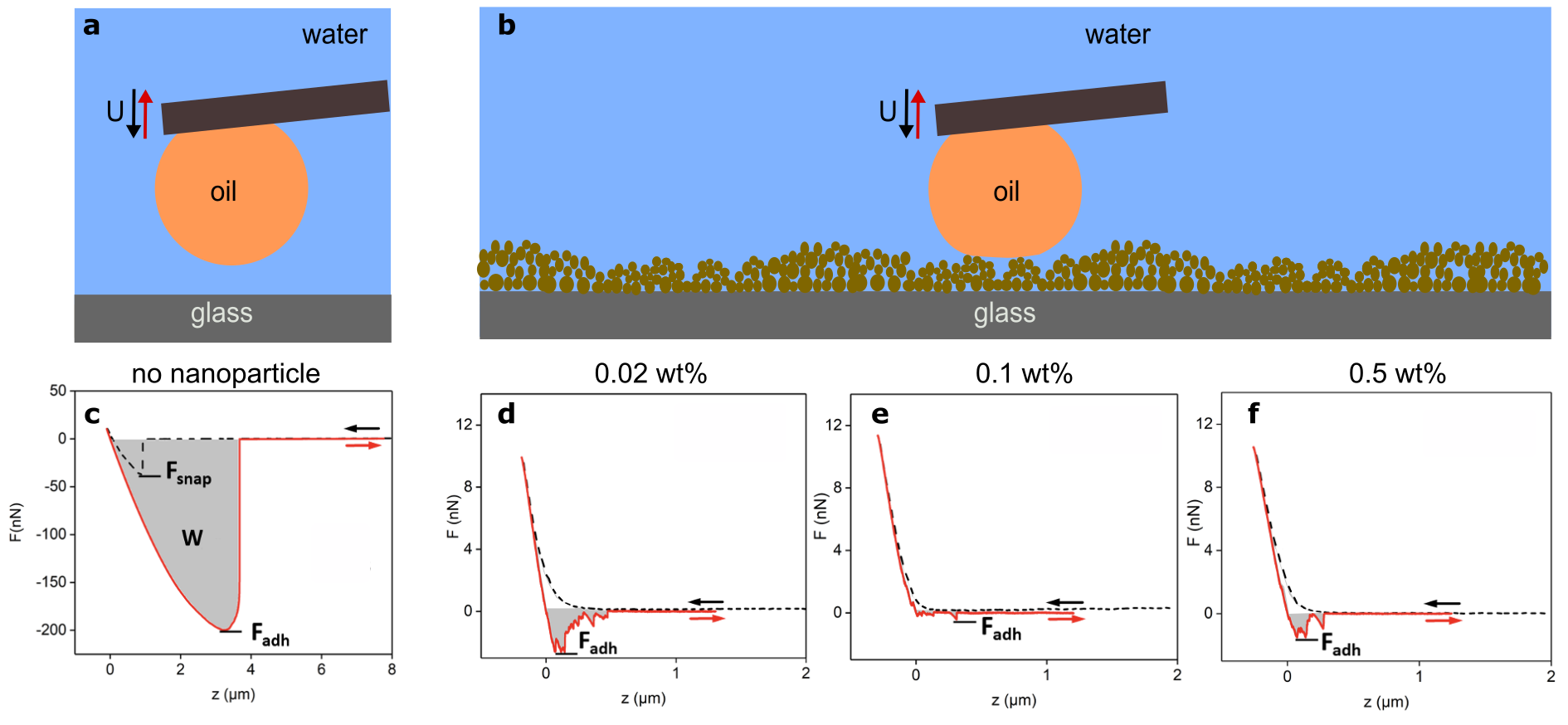}
\caption{\label{fig:dpafm} Droplet probe AFM can be used to quantify the interaction forces between an oil microdroplet and the surface. (a, b) Schematic showing droplet probe AFM for surfaces without and with nanoparticles. (c) Adhesion force measurement curve for the case with no nanoparticles. (d--f) Adhesion force measurement curves for the surface treated with 0.02-0.5 wt\% nanoparticles.}
\end{figure*}

 We attached a silicone oil droplet (diameter $2R$ = 20 $\mu$m) onto a tipless cantilever probe with a flexular spring constant of $k_z$ = 0.2 N m$^{-1}$ (Fig.~\ref{fig:dpafm}a, b). The force acting on the droplet $F$ as it approaches and retracts from the surface at a controlled speed $U$ = 2 $\mu$m s$^{-1}$ is obtained from the cantilever deflection $\delta_z$, since $F$ = $k_z \delta_z$ \cite{butt2005force}. We chose a relatively low $U$ to minimize viscous forces and hence observe the dynamics of contact line pinning-depinning.  
 
 Figure \ref{fig:dpafm}c shows the force spectroscopy measurements for a glass surface with no nanoparticles, with the approach and retract curves indicated by black dashed and red solid lines, respectively. When the droplet was far from the surface, the AFM did not detect any force, i.e. F = 0 (Fig.~\ref{fig:dpafm}c, dashed black line); however, upon contact, there was a sudden attractive snap-in force $F_{\text{snap}}$ = 39 nN. We continued to press onto the microdroplet to the maximum loading force of 10 nN, before retracting (Fig.~\ref{fig:dpafm}c, solid red line). For the droplet to be completely detached from the surface, there is a maximum adhesion force that must be overcome $F_\text{adh}$ = 210 nN. By integrating under the retract curve, we can also obtain the amount of work required to remove the droplet, $W$ = 0.6 pJ, which is a significant fraction of the total surface energy of the droplet $4 \pi R^{2} \gamma \approx$ 5 pJ where $\gamma$ is the surface tension, reflecting the fact that the oil droplet strongly adheres to glass in the absence of nanoparticles.
 
The results of force spectroscopy measurements are qualitatively different with the addition of nanoparticles (Fig.~\ref{fig:dpafm}d--f and Table \ref{tab:adhesion}). $F_{\text{snap}}$ becomes too small to measure and there is a significant reduction $F_\text{adh}$ and $W$ even at the smallest particle concentration used (0.02 wt$\%$) where there is no full coverage of nanoparticles. This is consistent with the stable water film observed using confocal RICM (Fig.~\ref{fig:diff_conc}a). The greatest reductions in $F_\text{adh}$ = 0.5 $\pm$ 0.3 nN  and $W$ = 0.8 $\pm$ 0.4 $\times$ 10$^{-16}$ J are obtained with 0.1 wt$\%$ particle concentrations (by more than 400 and 7000 times, respectively). However, the differences in $F_\text{adh}$ and $W$ are not significant as we vary the particle concentrations from 0.02--0.5 wt$\%$. Since adsorption of nanoparticles is a random process, there will be spatial variations in resulting micro-/nano-structures. The exact $F_\text{adh}$ and $W$ values will therefore depend on where the droplet lands on the surface. The errors reported for $F_\text{adh}$ in Table \ref{tab:adhesion} are not instrument errors, but rather reflect the spatial differences micro-/nano-structures. Similar reductions in $F_\text{adh}$ and $W$ were also observed as we vary the pH of the nanofluids from 2 to 5 (See Supplemental Figures 1 and 2). 
 
 We can estimate the contact radius of the droplet $r$ by noting that $F = (2 \gamma/R) \pi r^{2}$, where $\gamma = 50$ mN m$^{-1}$ is the water-oil interfacial tension as measured using the pendant drop's method and $2\gamma/R$ is the Laplace pressure inside the microdroplet. At the maximum loading $F$ = 10 nN, $r \approx$ 0.5 $\mu$m which is much larger than the size of the nanoparticles and similar in size to the spacing between nanoparticles aggregates. The droplet is therefore in contact with multiple nanoparticles and nanoparticle aggregates. This is in contrast to chemical force microscopy which uses a sharp nanometric sized solid tip to contact one or at most a handful of nanoparticles.
 
The nanoparticles and nanoparticle aggregates act as pinning points and droplet probe AFM can resolve the pinning-depinning dynamics as the droplet detaches from the surface \cite{daniel2019mapping}. In Figures \ref{fig:dpafm}d--f, we observe discontinuities in the retract curve with discrete force jumps $\delta F$ = 0.1--1 nN, which correspond to detachment from individual contact points. In contrast, the retract curve is smooth and continuous in the absence of nanoparticles, since the glass substrate is homogeneous (in the nanometre scale).   

\begin{table}[h!]
\caption{\label{tab:adhesion} Summary of the adhesion force and energy required to detach the oil microdroplet as measured using droplet probe AFM for different nanoparticle (NP) concentrations (See Fig.~\ref{fig:dpafm}c--f). The errors reported are the standard deviation for at least ten different spots.}. 
\begin{tabular}{ccc}
	NP conc. (wt$\%$) & $F_\text{adh}$ (nN) & $W$ (10$^{-16}$ J) \\
	\hline
	0 & 210 $\pm$ 10 & 6000 $\pm$ 1000 \\
	0.02 & 2.1 $\pm$ 0.6 & 4 $\pm$ 2   \\
	0.1  & 0.5 $\pm$ 0.3 & 0.8 $\pm$ 0.4 \\
	0.5 & 1.5 $\pm$ 0.7 & 2 $\pm$ 1 \\
\end{tabular}
\end{table}

\subsection*{Relevance to enhanced oil recovery application}

The wettability of reservoirs plays a critical role in enhanced oil recovery. Traditionally, wettability alteration is achieved through chemical surfactants, but they are usually expensive and do not perform well under high salinity conditions \cite{al2018wettability, liu2020review, dahbag2020suitability}. Here, we demonstrate how silica nanofluids can be used as effective alternative wetting agents even in the presence of salt.

So far we have shown wettability alteration on flat glass surfaces, which are different from rock surfaces with irregularly-shaped pores and crevices. We found analogous wettability alteration by silica nanofluids on a commercially available microfluidic chip or micromodel with etched glass channels that resemble a physical rock network. The micromodel has a height $H$ = 20 $\mu$m and a typical channel width $w$ $\sim$ 100 $\mu$m. The glass channels are initially hydrophobic but can be rendered hydrophilic by silica nanofluids \cite{ramakrishnaiah2016effect}.  

As described in our previous publications \cite{li2019investigation}, we first flowed in 0.2 wt$\%$ nanoparticle dispersion at flow rates of 5 $\mu$L min$^{-1}$ for 30 mins to alter wettability of micromodel, before flowing in the oil phase (limonene). To distinguish between water and oil, we have added two different fluorescent dyes: Rhodamine 6G in the aqueous phase (10 g L$^{-1}$, excitation and emission wavelengths $\lambda_{\text{ex, em}}$= 488, 525 nm) and Nile Red in the oil phase (5 g L$^{-1}$, $\lambda_{\text{ex, em}}$= 561, 635 nm).

\begin{figure}[!htb]
\centering
\includegraphics[scale=1.0]{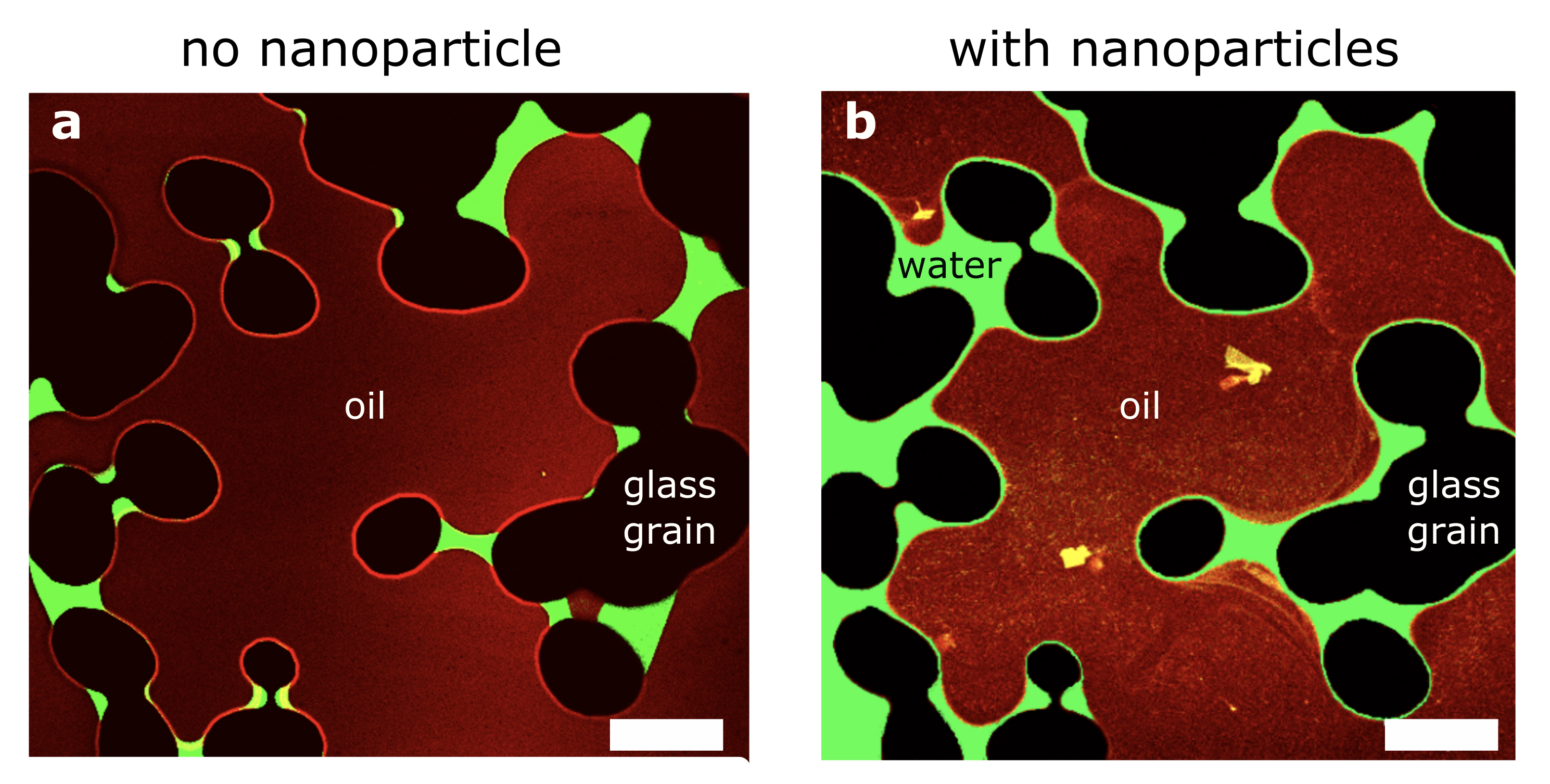}
\caption{\label{fig:fluor} Wetting states of the micromodel (a) without and (b) with nanoparticles. Scale bars are 100 $\mu$m }
\end{figure}

The resulting wetting states can then be observed under confocal fluorescence microscopy, with the aqueous and oil phase fluorescing green and red lights, respectively (Fig.~\ref{fig:fluor}). The glass grains inside the micromodel appear black because they do not contain any fluorescent dyes. In the absence of nanoparticles, while some water remains, the oil phase is directly in contact with the glass grains whose outlines appear red stained by the Nile Red dye (Fig.~\ref{fig:fluor}a). In contrast, the nanoparticles are able to stabilize a water film around the glass grains with no contact with the oil phase, resulting in more effective oil recovery as will be shown below.

\begin{figure}[!htb]
\centering
\includegraphics[scale=1.0]{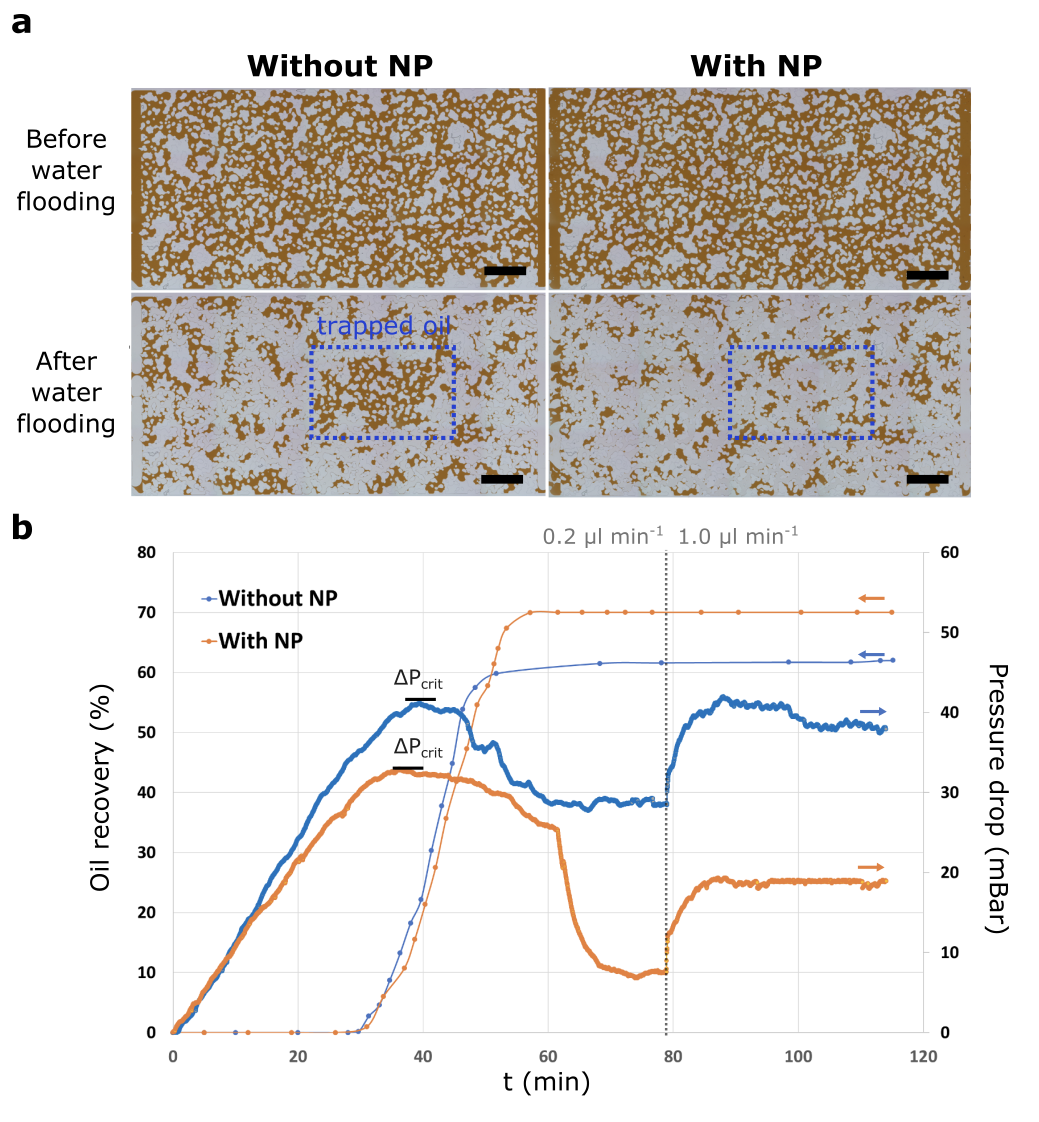}
\caption{\label{fig:flooding} Photographs showing the oil distribution inside microfluidic chips with/without nanoparticle (NP) treatment. Crude oil appears brown, while glass grains and water appear white. Scale bar is 2mm. (b) Oil recovery rates and pressure drop curves during water flooding with and without nanoparticles treatment.}
\end{figure}

To compare the effectiveness of crude oil recovery with and without nanoparticles, we flowed in 3 wt$\%$ brine solutions (similar salt content to seawater) at a flow rate of 0.2 $\mu$l min$^{-1}$ for 80 minutes before increasing it to 1.0 $\mu$l min$^{-1}$, i.e. water flooding \cite{li2019impact, li2018silica, li2015experimental}. Before water flooding, the microfluidic channels were filled with crude oil which appear brown (Fig.~\ref{fig:flooding}a). In the absence of nanoparticles, there was a large trapped oil bank in the middle of the micromodel even after water flooding. With the addition of the nanoparticles, more of the crude oil was removed including the previously trapped oil bank.

%The presence of this water film explains the effectiveness of fumed silica nanoparticles in enhanced oil recovery, as discussed in our previous publications \cite{li2019impact, li2018silica, li2015experimental}. 

We can quantify the oil recovery rates by tracking the area coverage of the crude oil in the microfluidic chip. After 1 hour of 0.2 $\mu$l min$^{-1}$ flow rate, we were able to recover 62$\%$ and 70$\%$ of the crude oil without and with nanoparticles, respectively, i.e. an improvement of 8$\%$ which is significant in EOR applications. There was no further improvement in recovery rates with longer flow times or increased flow rate of 1.0 $\mu$l min$^{-1}$.

We also tracked the pressure drop $\Delta P$ across the micromodel during the water flooding process. For oil recovery to happen, the pressure drop must overcome the capillary pressure and reach a critical pressure $\Delta P_{\text{crit}} \approx \gamma /(H + w) = 30$ mBar, where $\gamma$ = 50 mN m$^{-1}$ is the oil-water interfacial tension \cite{de2004capillarity}. The exact $\Delta P_{\text{crit}}$ depends on the shape of the water-oil meniscus, in particular the contact angle it makes on the glass grains at the point of displacement.  Experimentally we found that crude oil displacement started at $t$ = 30 min, with $\Delta P_{\text{crit}}$ = 32 mBar with nanoparticles lower than $\Delta P_{\text{crit}}$ = 42 mBar without nanoparticles. Beyond this, the pressure starts to fall and $\Delta P$ is consistently lower with nanoparticle treatment.

\section*{Conclusions}

In conclusion, we have demonstrated that fumed silica nanoparticles can form a porous layer which stabilizes a thin water film, which we can directly visualize using RICM. This greatly reduces the adhesion of oil droplets even if the surface is only partially covered by nanoparticles. Using droplet probe AFM, we are able to directly measure the interactions of an oil microdroplet with the adsorbed nanoparticles, including contact line pinning-depinning dynamics. We also showed how wettability alteration by silica nanoparticles can improve the oil recovery rate by 8$\%$ in a model microfluidic system.  

\section*{Materials and Methods}

\textbf{Materials and chemicals.} Hydrophilic fumed silica nanoparticles were obtained from Evonik Industries. They have an average primary particle size of 7 nm and specific surface area of 300 m$^{2}$ g$^{-1}$. Microscope slides were purchased from AmScope. Limonene was purchased from Fisher Scientific, while Rhodamine 6G dye, Nile Red dyes, sodium chloride (NaCl) and silicone oil (Polyphenyl-methylsiloxane with viscosity around 100 mPa.s and density 1.06 g/ml) were purchased from Sigma-Aldrich. Reprorubber $\textregistered$ Thin Pour silicone elastomer (two-part cure) was purchased from Flexbar. All chemicals were used as received. Deionized (DI) water with a resistivity of 17 M$\Omega$ cm was used as the aqueous phase. The crude oil has a density of 0.89 g/ml and viscosity of 41 mPa s. 

The micromodel was purchased from Micronit Micro Technologies. The pore structure in the microfluidic chip, based on the structure of a physical rock, is etched with hydrofluoric acid and the surface is initially hydrophobic. The microfluidic channels have a height $H$ = 20 $\mu$m and a permeability of 2.5 Darcy. Pore structure and parameters of the micromodel can be found in our previous publication \cite{li2019investigation}.
 
\textbf{Preparation of nanofluids.} Three nanoparticle concentrations (0.02, 0.1 and 0.5 wt$\%$) were dispersed in 3 wt$\%$ NaCl aqueous solution (brine) using an ultrasonicator (UP400St, Hielscher, Germany). The pH of nanofluid was adjusted by adding concentrated hydrochloric acid (HCl). Note that the pH of DI water is lowered by the addition of nanoparticles which have a large surface-to-volume ratio (300 m$^{2}$ g$^{-1}$). For example, a 1 wt$\%$ nanoparticle concentration lowers the pH to 4.8 even without HCl.

\textbf{Wettability alteration of glass surfaces.} Glass substrates (2 cm by 2 cm) were cut to size from microscope slides and soaked in nanofluid/brine solution for 1 hour. The glass substrates with the adsorbed nanoparticle layer were then rinsed (to remove any loose nanoparticles) and  transferred to a petri-dish of brine water for further experiments, e.g. droplet probe AFM.

\textbf{Cryo-SEM imaging}.
Prior to cryo-preparation, each sample was immersed in nanofluids with different concentrations for one hour and then rinsed with deionized water to remove the salt and free nanoparticles. Next, the samples were cooled by plunging into a nitrogen slush at atmospheric pressure. Then the samples were frozen at -120$^{\circ}$C, etched for 300 s at -95$^{\circ}$C, and sputtered with gold in the cryo-preparation chamber. After preparation, samples were transferred into the SEM chamber (JSM-6700F, JEOL, Japan) for imaging. 

\textbf{AFM topography measurements.} Before exposing the glass surface to the nanofluid, a portion of the surface was covered with Reprorubber $\textregistered$  silicone elastomer. After soaking the surface in nanofluid for about an hour, the silicone elastomer was peeled off to reveal the underlying glass substrate. AFM topography (JPK Instrument) was performed under water at the boundary between the bare glass surface (previously covered by elastomer) and the nanoparticle layer.

\textbf{Confocal RICM.} We first deposited a millimetric-sized oil droplet on the substrate of interest under water. RICM was performed using a confocal laser scanning microscope (FV3000RS; Olympus Corporation). During the experiment we raster scanned the surface with one focused beam of monochromatic light with wavelength $\gamma$=561 nm, and captured the reflected light through the pinhole of a confocal microscope; thus, only reflected light from the focal plane, that was, the interface of interest, was able to reach the photomultiplier tube of the microscope \cite{daniel2017oleoplaning}. In the presence of a thin water film, the light reflected off the solid/water and water/oil droplet interfaces would interfere with one another constructively or destructively to give bright or dark fringes, respectively.

\textbf{Droplet probe AFM.} To create small oil droplets, we force 60 $\mu$l of oil through a small capillary tube with inner and outer diameters of 360 $\mu$m and 290 $\mu$m into a petri-dish of water. This generates multiple droplets with a broad range of sizes $2R$ = 20---200 $\mu$m. We can then pick up a droplet of the desired size (approximately 20 $\mu$m) using a tipless silicon cantilever, with dimensions width $w$ = 50 $\mu$m and length $l$ = 450 $\mu$m, and a spring constant $k$ = 0.2 N m$^{-1}$ as determined using Sader's method. Once picked up, the droplet adheres much more strongly to cantilever than to the surface, allowing us to perform force spectroscopy measurements on the surface of interest multiple times without detaching. For each surface, we performed force spectroscopy measurements for at least 10 different spots, with 3--5 repeats on each spot. During each force spectroscopy measurement, the droplet approaches and retracts from the surface at a controlled speed $U$ = 2 $\mu$m s$^{-1}$, with the force experienced by the droplet inferred from the cantilever deflection $\delta_{z}$, since $F = k \, \delta_{z}$. $\delta_{z}$ is detected by shining a laser light (infrared, wavelength of 980 nm) onto the cantilever, which is reflected into a 4-quadrant sensor. More details can be found in our previous publication \cite{daniel2019mapping}. 

\textbf{Microfluidic flooding experiment.} The micromodels were first completely filled with 3 wt$\%$ NaCl brine solution. To alter the wettability of the micromodel, we injected 1 wt$\%$ silica nanofluids for 1 hour at 1 $\mu$l min$^{-1}$ flow rate; for the control, no nanofluid was injected. The crude oil was then injected at 1 $\mu$l min$^{-1}$ flow rate for 1 hour. To recover the crude oil, 3 wt$\%$ NaCl brine solution was again injected first at 0.2 $\mu$l min$^{-1}$ flow rate for 80 min, before increasing it to 1.0 $\mu$l min$^{-1}$

Local images of the micromodel were taken using Leica digital microscope DVM6 and then stitched to get a whole picture of the micromodel. ImageJ was used to perform image analysis for oil recovery calculations. Pressure drop across the micromodel was recorded during water flooding.

\begin{acknowledgement}
This research was supported by the Agency for Science, Technology and Research (A*STAR), Singapore, under IAFPP Programme (project title: Advanced Functional Polymer Particle Technologies for the Oil and Gas Industry; grant no.: A18B4a0094) as well as the Petroleum Engineering Professorship Grant from the Economic Development Board of Singapore. The authors would like to thank Ms. Jing Lin Inez Kwek for her help with Cryo-SEM imaging.
\end{acknowledgement}

%%%%%%%%%%%%%%%%%%%%%%%%%%%%%%%%%%%%%%%%%%%%%%%%%%%%%%%%%%%%%%%%%%%%%
%% The same is true for Supplemental Information, which should use the
\begin{suppinfo}
%
%This will usually read something like: ``Experimental procedures and
%characterization data for all new compounds. The class will
%automatically add a sentence pointing to the information on-line:
%

Additional experimental data can be found in the Supporting Information.
\end{suppinfo}

%\bibliography{biblio}

\providecommand{\noopsort}[1]{}\providecommand{\singleletter}[1]{#1}%
\providecommand{\latin}[1]{#1}
\makeatletter
\providecommand{\doi}
  {\begingroup\let\do\@makeother\dospecials
  \catcode`\{=1 \catcode`\}=2 \doi@aux}
\providecommand{\doi@aux}[1]{\endgroup\texttt{#1}}
\makeatother
\providecommand*\mcitethebibliography{\thebibliography}
\csname @ifundefined\endcsname{endmcitethebibliography}
  {\let\endmcitethebibliography\endthebibliography}{}

\end{document}